\documentclass[12pt, a4paper]{article}

\usepackage[table]{xcolor}
\usepackage{amsmath,amssymb,amsfonts,amsthm}
\usepackage[pdftex]{graphicx}
\usepackage{bm}
\usepackage{booktabs}

\usepackage{float}

\title{\bf Formulations to select assets for constructing sparse index tracking portfolios}
\author{
\normalsize Yutaka Sakurai
\thanks{\ \ \scriptsize RP Tech., AI Finance Application Research Institute, 
Placeo Aoyama Building, 3rd Floor 2-7-13 Kitaaoyama Minato-ku, Tokyo 107-0061, Japan}
\\ \normalsize ysakurai@rptech.co.jp
\and \normalsize Daiki Wakabayashi \thanks{\ \ \scriptsize FactSet Pacific Inc., 
Otemachi One Tower 6F, 1-2-1 Otemachi Chiyoda-ku,
Tokyo 100-0004 Japan}  
\\ \normalsize dwakabayashi@factset.com
\and \normalsize Fumio Ishizaki \footnotemark[1] \thanks{\ \ \scriptsize Modal Stage Inc.,
4-7-6-704 Shiba, Minato-ku, Tokyo 108-0014, Japan
}  
\\ \normalsize ishizaki@modalstage.com
}

\date{}
\begin{document}
\maketitle
\begin{abstract}
\noindent 
In this paper, we study asset selection methods to construct a sparse index tracking portfolio. 
For its advantage over full replication portfolio, 
the concept of sparse index tracking portfolio has significant attention in the field of 
finance and investment management. 
We propose useful formulations to select assets for sparse index tracking portfolio. 
Our formulations are described as combinatorial optimization problems, 
and they can yield various asset selection methods,  including some existing methods, 
by adjusting the values of parameters.  
As a result, the proposed formulations can provide a well-balanced asset selection 
to create successful sparse index tracking portfolios. 
We also provide numerical examples to compare 
the tracking performance of resulting sparse index tracking portfolios.  

{\flushleft{{\bf KEYWORDS:} sparse index tracking portfolio; combinatorial optimization problem; correlation distance; asset selection with multi-stage.}}
\end{abstract}

\section{Introduction}
\label{sec:Introduction}

A stock index is composed of a collection of stocks, which captures the value of its specific stock market. 
A stock index is a hypothetical portfolio of assets in the sense that 
we cannot directly invest on it, i.e., a stock index is not a financial instrument that we can trade \cite{benidis2018optimization}. 
Thus, in order to gain access to a stock index, we need to create a portfolio of assets that tracks closely a given stock index. 
The most straightforward manner to create a tracking portfolio of stock index 
is to buy all the assets composing the stock index with weights that are identical to their respective weights in the stock index. 
This approach is known as full replication, where a perfect tracking can be achieved. 
However, the full replication approach has several drawbacks in practice \cite{benidis2018optimization}. 
First, the execution of such a full replication portfolio may be complicated and expensive, because it can consist of thousands of stocks.
Second, a portfolio consisting of all the assets usually includes many small and illiquid stocks. 
This imposes higher risk on investors, because an illiquid stock is hard to buy/sell when they are looking to entry/exit 
and it increases the costs due to slippage. 
Furthermore, since every asset is dealt with a separate transaction, a full replication portfolio leads to significantly 
large commission fees. 

A practical approach to overcome those drawbacks of full replication approach is to use a small number of assets to approximately replicate 
a stock index. 
In this approach, we construct a {\it sparse index tracking portfolio} (see, e.g, \cite{beasley2003evolutionary,jansen2002optimal,varsei2013heuristic} and references therein), 
which can closely track the index using a small number of assets. 
A sparse portfolio can simplify the execution of the tracking portfolio and avoid handling illiquid stocks. 
Furthermore, since only a small number of assets is dealt, the transaction fees are significantly reduced.   
For those merits, 
the concept of sparse index tracking portfolio has significant attention in the field of 
finance and investment management. 

The approaches to constructing a sparse index tracking portfolio can be categorized into the two types of approaches: 
Two-step approach and Joint approach \cite{benidis2018optimization,feng2016signal}. 
The two-step approach of sparse index tracking portfolio decomposes the task into two steps: 
(1) Selection of assets and (2) Determination of their weights. 
More precisely, in the first step, 
we select a subset of assets among all the assets composing the stock index. 
In the second step, we determine the weights of the selected assets in the sparse index tracking portfolio. 
The joint approach of sparse index tracking portfolio unifies the two steps of the two-step approach and penalizes the cardinality of 
the tracking portfolio with regularization \cite{benidis2017sparse,fastrich2015constructing,feng2016signal}. 
Since the joint approach implicitly determines the number of selected assets by regularization, it is not easy to 
control the number of selected assets, compared to the two-step approach,  although it is clear how optimal the resulting 
tracking portfolio is. 
On the other hand, in the two-step approach, the number of selected assets is usually incorporated explicitly as a constraint 
in optimization problem.

In this paper,  we study the two-step approach for sparse index tracking portfolio.  
In particular, we focus on the first step of the two-step approach and 
propose useful formulations to select assets for sparse index tracking portfolio. 

As the first step of the two-step approach, 
several asset selection methods have been proposed. 
A simple and naive method is to randomly select a predetermined number $M$ of assets among all the assets composing stock index 
\cite{benidis2018optimization,dose2005clustering}.
This random selection is mainly used as a benchmark.  
The most widely used selection method is to select a predetermined number $M$ of  the largest assets 
according to their market capitalizations (MCs) \cite{benidis2018optimization}. 
We hereafter call this selection method the MC top-tier selection method.   
The MC top-tier selection does not consider the correlation among the movements of assets composing the stock index. 
This could potentially be a factor leading to a degradation in its tracking performance. 
Recently \cite{hong2021market} has proposed an asset selection method formulated in the form of 
a binary quadratic $K$-medoid graph-clustering problem \cite{bauckhage2019qubo} or more generally 
combinatorial optimization problem. 
We hereafter call this selection method the correlation balance selection (CBS) method. 
Although 
CBS considers the correlation among the movements of assets composing the stock index, 
it ignores the impact of assets  with large MC on the stock index. 
This could potentially be a factor leading to a serious degradation in its tracking performance. 
In this paper, in order to overcome those drawbacks of both MC top-tier selection method and CBS method, 
we propose formulations for selection method,  which integrate them 
and then include them as special cases. 
Our formulations can yield various mixture methods of those two methods by adjusting (hyper) parameters in 
the formulations and provide a well-balanced asset selection suppressing their drawbacks.

The remainder of this paper is organized as follows. 
Section 2 proposes formulations to select assets for sparse index tracking portfolio.
We represent a single-stage selection method and a multi-stage selection method in Section 2. 
Our formulations proposed in Section 2 are described as combinatorial optimization problems. 
In recent years, there has been a great interest in combinatorial optimization problems 
in the area of investment strategies 
(see, e.g., \cite{alipour2016quantum,hong2021market,rosenberg2016solving,sakurai2021correlation,tatsumura2023real} and reference therein), 
according to the development of heuristics and solvers for combinatorial optimization problems. 
In Section 3, we apply the proposed formulations to construct tracking portfolios of TOPIX (Tokyo Stock Price Index)
and those of S\&P500 (Standard \& Poor's 500),  
and we provide some numerical examples to compare the tracking performance of the resulting tracking portfolios.  
Conclusion is drawn in Section 4.

\section{Formulations to select assets}
\label{sec:Formulation}

In this section, we provide formulations to select assets for sparse index tracking portfolios. 
We suppose a stock index which is composed of $L$ assets and weighted by their MCs. 
We then  consider constructing sparse index tracking portfolios which track the stock index and 
are composed of $M$ assets $(M < L)$. 
%
    
Without loss of generality, we assume that 
the $L$ assets composing of stock index are sorted in descending order of their MCs. 
We index the $L$ assets by the descending order and hereafter call  an asset located on the $l$th position in the order the $l$th asset.  
We consider only the top $K$ assets $(K \leq L)$ in the following formulations. 

We first define correlation distance $d_{ij}$ $(i, j = 1, \ldots, K)$ between the $i$th assets and the $j$th asset  by 
\begin{equation}
d_{ij}  = \sqrt{2(1-\rho_{ij})}, 
\label{eq:1}
\end{equation}
where $\rho_{ij}$ denotes the correlation coefficient between 
the return of the $i$th asset and that of the $j$th asset.  
It is known that the correlation distance $d_{ij}$ fulfills all three properties that 
must be satisfied by a metric distance \cite{stanley2000introduction}. 
Let $\bm{D}$ denote the $K \times K$ correlation distance matrix whose $(i,j)$th element is given  by $d_{ij}$.

We always select the top $N$ $(0 \leq N \leq M)$ MC assets. 
Also, we impose a restriction that selected stocks must be within the top $H$ $(N \leq M \leq H \leq K)$ MC.  
This ensures that selected stocks have a certain level of liquidity. 
We then define $x_{k}=1$ for $k=1, \ldots, N$, 
$x_{k} = 0$ for $k=H+1, \ldots, K$ 
and  
for $k=N+1, \ldots, H$, 
\begin{equation}
x_{k}=
\begin{cases}
1 & (\mbox{if the $k$th asset is selected for the tracking portfolio}), \\
0 & (\mbox{otherwise}).
\end{cases}
\end{equation}
Note here that among $x_{k}$ $(k=1, \ldots, K)$, only $x_{N+1}, \ldots, x_{H}$ are decision variables  
and the number of them is equal to $H-N$.  
For notational convenience, 
we introduce  a column vector $\bm{x}  = (x_{1}, \ldots, x_{N}, x_{N+1}, \ldots, x_{H}, x_{H+1}, \ldots, x_{K})^\mathsf{T}$. 

We then consider the following quadratic binary optimization problem:  
\begin{equation}
\begin{aligned}
& {\text{min}}
& & \beta \bm{x}^\mathsf{T} \bm{D} \bm{e} - \frac{1}{2} \alpha \bm{x}^\mathsf{T} \bm{D} \bm{x}, \\
& \text{subject to}
& & \bm{x}^\mathsf{T} \bm{e} = M, 
\end{aligned}
\label{eq:MP}
\end{equation}
where $\alpha$ and $\beta$ are hyper parameters with nonnegative real values, and 
$\bm{e}$ denotes the $K \times 1$ vector whose elements are all equal to one.  
The first term of the objective function in (\ref{eq:MP}) acts to 
select assets with small sum of correlation distances to other assets.  
In other words, it aims to select assets that are positioned as centrally as possible among all the candidate assets. 
The second term of the objective function in (\ref{eq:MP}) acts to make selected assets as far apart as possible. 
In other words, it aims to ensure that assets to be selected have dissimilar behavior with each other. 
To construct a successful tracking portfolio, it is crucial to take a balance between those {\it centrality} and {\it dissimilarity} elements 
upon selecting assets. 
The hyper parameters $\beta$ and $\alpha$ govern centrality and dissimilarity, respectively, 
and they are designed to adjust the balance between centrality and dissimilarity elements. 
More precisely, when $\alpha$ is fixed and $\beta$ is increased, assets that are centrally located are more likely to be selected, 
and when $\beta$ is fixed and $\alpha$ is increased, assets that are dissimilar to each other are more likely to be selected. 
We hereafter call $\beta$ and $\alpha$ the centrality parameter and the dissimilarity parameter, respectively. 
The constraint in (\ref{eq:MP}) ensures that the number of selected assets is equal to $M$. 
From the solution $\bm{x}$ of (\ref{eq:MP}), the set of the selected assets is then described as 
\begin{equation}
{\cal K}  = \{ k | x_{k} = 1 \mbox{ for } k=1, \ldots, H \}.  
\label{eq:SA}
\end{equation}

Note here that when we set $M=N$ in our formulation (\ref{eq:MP}), 
we perform MC top-tier selection method for selecting assets \cite{benidis2018optimization}. 
In this case, the selected assets are independent of both hyper parameters $\alpha$ and $\beta$. 
Also,  note here that when we set $N=0$ in our formulation (\ref{eq:MP}), 
we perform CBS method for selecting assets \cite{hong2021market}.   
Our formulation therefore includes both MC top-tier selection and CBS methods as special cases, 
and it can yield various mixture methods of these methods by adjusting (hyper) parameters $M$, $N$, $\alpha$ and $\beta$. 
The utilization of our formulation may provide a well-balanced asset selection that overcomes the shortcomings of those two methods.

\subsection{Extension to multi-stage selection}

To select assets for tracking portfolio, 
we have proposed  the formulation described as the quadratic binary optimization problem 
(\ref{eq:MP}) and (\ref{eq:SA}). 
The quadratic binary optimization problem  (\ref{eq:MP}) includes the dissimilarity parameter $\alpha$ and the centrality parameter $\beta$. 
It might be beneficial to select assets in multiple stages by adjusting the values of the parameters $\alpha$ and $\beta$. 
To accommodate this approach, we extend the single-stage selection introduced in this section  to a multi-stage selection.
We then consider the following quadratic binary optimization problem for multi-stage selection: 
\begin{equation}
\begin{aligned}
& {\text{min}}
& & \beta^{(i)} {\bm{x}^{(i)}}^\mathsf{T} \bm{D} \bm{e} 
- \frac{1}{2} \alpha^{(i)} {\bm{x}^{(i)}}^\mathsf{T} \bm{D} {\bm{x}^{(i)}}, \\
& \text{subject to}
& & {\bm{x}^{(i)}}^\mathsf{T} \bm{e} = M^{(i)}, 
\end{aligned}
\label{eq:MP2}
\end{equation}
where $\alpha^{(i)}$ and $\beta^{(i)}$ are hyper parameters with nonnegative real values, and 
$M^{(i)}$ is a hyper parameter. 
In other words, 
we repeatedly solve the quadratic binary optimization problem (\ref{eq:MP}) 
with different values of hyper parameters $\alpha$, $\beta$ and $M$.  

From the solution ${\bm{x}}^{(i)}$ of (\ref{eq:MP2}), the set of the selected assets is then described as 
\begin{equation}
{\cal K}^{(i)}  = \{ k | x^{(i)}_{k} = 1 \mbox{ for } k=1, \ldots, H \}.  
\label{eq:SA2}
\end{equation}
We consider the union ${\cal K}$ of ${\cal K}^{(i)}$, where ${\cal K}^{(i)}$ is the set of selected assets in the $i$th stage selection as follows: 
\begin{equation}
{\cal K} = \bigcup_{i} {\cal K}^{(i)}. 
\end{equation}
 Finally we adjust the number of selected assets by considering a subset $K^{*}$ of ${\cal K}$. 
 More specifically,  we obtain the set ${\cal K}^{*}$ of selected assets through multi-stage selection by 
\begin{equation}
{\cal K}^{*} =  \{ J | J \in {\cal K} \mbox{ and } \sum_{j=1}^{J} I(j \in {\cal K}) \leq M^{*} \}, 
\end{equation}
where  $I(j \in {\cal K})$ is defined by 
\begin{equation}
I(j \in {\cal K}) = 
\left\{ 
\begin{array}{cc}
1 & (j \in {\cal K}),  \\
0 & (j \not\in {\cal K}), \\
\end{array}
\right. 
\label{eq:9}
\end{equation}
and  $M^{*}$ is a hyper parameter which limits the number of selected assets to at most $M^{*}$.

\section{Numerical examples}

In this section, we provide some numerical examples in order to  
observe 
how the hyper parameter settings in our formulation affect tracking performance. 
We apply formulations proposed in Section 2 to 
construct sparse tracking portfolios of  TOPIX (Tokyo Stock Price Index) and 
those of S\&P500 (Standard \& Poor's 500). 
In the numerical examples, the period is set from Apr.\ 1, 2005 to Jan.\ 31 2023 (4366 business days) 
for TOPIX and from Jan.\ 2, 2014 to Jan.\ 31, 2024 (2537 business days) for S\&P500.

\begin{table}
  \begin{center}
  \caption{Parameters in Qontigo Axioma Portfolio Optimizer(TOPIX)}
  \label{Qontigo-TOPIX}
  \begin{tabular}{|l|l|} 
  \hline
  Risk Model & Axioma Japan Statistical Equity Risk Model MH 4 \\ \hline 
  Object & Minimize Tracking Error \\ \hline 
  Benchmark & TOPIX \\ \hline 
  Initial Portfolio & Cash \\ \hline 
  Trade Universe & Based on 20 Stock list \\ \hline 
  Constraints & Budget 10M JPY \\ \hline 
  Rebalance Timing & Every Calendar Quarter (3/6/9/12) end \\ 
  \hline
  \end{tabular}
  
  \end{center}
\end{table}
  
\begin{table}
  \begin{center}
  \caption{Parameters in Qontigo Axioma Portfolio Optimizer(S\&P500)}
  \label{Qontigo-S&P500}
  \begin{tabular}{|l|l|} 
  \hline
    Risk Model & Axioma US Fundamental Equity Risk Model MH 4 \\ \hline 
    Object & Minimize Active Risk \\ \hline 
    Benchmark & S\&P500\\ \hline 
    Initial Portfolio & Cash \\ \hline 
    Trade Universe & Based on 20 Stock list \\ \hline 
    Constraints & Budget 10M JPY equivalent \\ \hline 
    Rebalance Timing & Every Calendar Quarter (3/6/9/12) end \\ 
  \hline
  \end{tabular}
  \end{center}
\end{table}

To construct the tracking portfolio, 
we need any method to determine the weights of selected assets,  
because our formulations are just to select assets. 
For this purpose, 
we utilize  the optimization tool Qontigo Axioma Portfolio Optimizer, 
which is one of standard tools in the asset management industry. 
Tables 1 and 2 display parameters used in Qontigo Axioma Portfolio Optimizer 
for tracking portfolios of TOPIX and those for tracking portfolios of S\&P500, respectively.

\begin{table}
  \begin{center}
  \caption{Hyper parameter setting (TOPIX, S\&P500)}
  \label{hyper-parameter}
  \begin{tabular}{lcccccccc} 
  \toprule 
    Tracking portfolio & $N$ & \multicolumn{3}{c}{$M$} & \multicolumn{2}{c}{$\alpha$} & \multicolumn{2}{c}{$\beta$}  \\
  \midrule 
    E1 (MC top-tier) & 30 & \multicolumn{3}{c}{30} & \multicolumn{2}{c}{--} & \multicolumn{2}{c}{--} \\ 
    E2 &  10 & \multicolumn{3}{c}{30} &  \multicolumn{2}{c}{$\frac{1}{M}$} & \multicolumn{2}{c}{$\frac{1}{H}$} \\
    E3 &  5 & \multicolumn{3}{c}{30} &  \multicolumn{2}{c}{$\frac{1}{M}$} & \multicolumn{2}{c}{$\frac{1}{H}$} \\
    E4 (CBS with single-stage selection) &  0 & \multicolumn{3}{c}{30} &  \multicolumn{2}{c}{$\frac{1}{M}$} & \multicolumn{2}{c}{$\frac{1}{H}$} \\
  \toprule
    & $N$ & $M^{(1)}$ & $M^{(2)}$ & $M^{*}$ & $\alpha^{(1)}$ & $\alpha^{(2)}$ & $\beta^{(1)}$ & $\beta^{(2)}$  \\
  \midrule
    E5  (CBS with two-stage selection) & 0 & 20 & 20 & 30 & $\frac{1}{M^{(1)}}$ & $\frac{2}{M^{(2)}}$ &  $\frac{1}{H}$ & $\frac{1}{H}$ \\ 
    E6 &  5 & 20 & 20 & 30 & $\frac{1}{M^{(1)}}$ & $\frac{2}{M^{(2)}}$ &  $\frac{1}{H}$ & $\frac{1}{H}$ \\ 
  \bottomrule 
  \end{tabular}
    
  \end{center}
\end{table}

In the numerical examples, 
we consider six sparse tracking portfolios E1--E6, where E1--E4 is 
yielding by the one-stage selection  (\ref{eq:MP}) and (\ref{eq:SA}) 
and E5--E6 is yielded by  the two-stage selection (\ref{eq:MP2})--(\ref{eq:9}).  
Table \ref{hyper-parameter} 
displays their hyper parameter setting commonly used in our formulations 
for TOPIX and S\&P500.  
In the numerical examples, 
we consider selecting 30 stocks from the top 500 MC stocks. 
In other words, we set $K=500$ and $M=M^{*}=30$.
We also set $H=150$ in order to ensure liquidity and tradability of the stocks to be selected. 
Note here that 
E1 and E4 corresponds to MC top-tier portfolio and CBS portfolio with one-stage selection, respectively.  
Also note that E5 corresponds to CBS portfolio with two-stage selection. 
The correlation coefficient $\rho_{i,j}$ appearing in (\ref{eq:1}) and correlation distance 
$\bm{D} = [d_{i,j}]$ appearing 
in (\ref{eq:MP}) or (\ref{eq:MP2}) 
are calculated from the weekly logarithmic rate over the previous 
five years from the beginning of each period. In calculating the correlation coefficient, we use a shrinkage estimator 
and a linearly weighted sample. 

As shown in Table \ref{hyper-parameter}, we commonly set the dissimilarity parameter $\alpha$ as $\alpha = 1/M$ and 
the centrality parameter $\beta$ as $\beta = 1/H$ for E2--E6 in both cases of TOPIX and S\&P500.
The idea behind setting $\alpha = 1/M$ is a simple concept to make the second term of the objective function 
(\ref{eq:MP}) or (\ref{eq:MP2}), which is related to dissimilarity,  the average of the sum of distances between selected $M$ assets.
Also, the idea behind setting $\beta = 1/H$ is as follows. 
By considering $H$ candidate assets for selection, 
we attempt to make the first term of the objective function (\ref{eq:MP}) or (\ref{eq:MP2}), 
which is related to centrality, smaller. 
Setting $\beta = 1/H$ is a simple concept to make the first term the average of the total distance 
to the other stocks per candidate assets. 
While we may be able to fit the parameters $\alpha$ and $\beta$ to the data, 
considering the risk of overfitting, we decided to set as $\alpha = 1/M$ and $\beta = 1/H$ 
in the numerical examples.

Our formulations (\ref{eq:MP}) and (\ref{eq:MP2}) are  
combinatorial optimization problems,  which belong 
to the class of NP-hard in terms of computational complexity. 
In the numerical examples, we first utilize Simulated Annealing as heuristics 
to find an approximate solution, and then apply a post-processing to the approximate solution 
to improve the quality of solution. 
In the post-processing, we first select a pair of $(x_{i}, x_{j})$ such that $x_{i} = 1$ and $x_{j}=0$ and 
swap their values as $x_{i} = 0$ and $x_{j}=1$. If and only if the objective function decreases by swapping their values, 
we actually swap their values as $x_{i} = 0$ and $x_{j}=1$. 
We repeat this process for all asset pairs in the post-processing.


%

In what follows, we examine the tracking performance of the tracking portfolios E1--E6 
for TOPIX. 
Let $R_{t}$  and $\hat{R}_{t}$ $(t=1,2,\ldots, 4366)$ denote 
the daily return of stock index at day $t$  and 
the daily return of tracking portfolio at day $t$, respectively. 
We then define the cumulative return $C^{p}_{t}$ of the stock index 
from  day $t$ to day $t+p-1$ with time horizon $p$ $(p=1,2,\ldots)$  
and 
the cumulative return $\hat{C}^{p}_{t}$ of the resulting tracking portfolio 
from day $t$  to day $t+p-1$ with time horizon $p$ as 
\begin{equation}
C^{p}_{t} 
= \prod_{n=t}^{t+p-1} (1 + R_{n}) -1, 
\qquad 
\hat{C}^{p}_{t} 
= \prod_{n=t}^{t+p-1} (1+ \hat{R}_{n}) -1. 
\end{equation}
Note here that when the time horizon $p$ is set as $p=1$, 
the cumulative returns $C^{p}_{t}$ and $\hat{C}^{p}_{t}$ are 
reduced to the daily returns $R_{t}$ and $\hat{R}_{t}$, 
respectively.
We then investigate the residual of cumulative return 
$\varepsilon^{p}_{t} = \hat{C}^{p}_{t} - C^{p}_{t}$ for time horizon $p$ 
as an estimate 
for the tracking error of the tracking portfolios. 
More specifically, 
we statistically examine the bias and variance of residuals $\varepsilon^{p}_{t}$ 
for time horizon $p=1, 10, 50, 100$, where we set the sample size to 200 
through all the statistical tests provided in this section.  

\begin{table}
  \begin{center}
    \caption{Sample mean of residual for cumulative return (TOPIX) ($\times 10^{-5}$)}
    \label{sample-mean-TOPIX}
    \begin{tabular}{crrrrrr} 
    \toprule 
     & E1 & E2 & E3 & E4 & E5 & E6 \\
    \midrule 
    $p=1$ & $-0.746$ & $0.347$ & $0.931$ & $-2.68$ 
    & $0.269$ & $1.29$ \\
    $p=10$ & $-6.56$ & $0.442$ & $6.16$ & $-30.4$
    & $-2.33$ & $8.25$ \\
    $p=50$ & $-49.7$ & $-26.6$ & $13.4$ & $\cellcolor{blue!10} \textcolor{gray}{-156}$    
    & $-8.37$ & $20.8$ \\
    $p=100$ & $-110$ 
    & $-57.7$ & $35.8$  
    & $\cellcolor{blue!10} \textcolor{gray}{-283}$
    & $15.1$ & $58.9$ \\
    \bottomrule 
    \end{tabular}
      
  \end{center}
  \end{table}

We first focus on the biases of residuals $\varepsilon^{p}_{t}$ 
as estimates for the tracking error of the tracking portfolios E1--E6.
Before examining the biases of residuals, 
we check the normality of residuals $\varepsilon^{p}_{t}$ of E1--E6. 
For all the cases of $p=1,10,50, 100$, 
the Shapiro-Wilk test indicated that the residuals $\varepsilon^{p}_{t}$ of E1--E6 do not all follow 
normal distributions at the 5\% significance level. 
To examine the biases of the residuals, 
we thus apply the Wilcoxon signed-rank test, which is a non-parametric rank test, 
to the residuals $\varepsilon^{p}_{t}$ of E1--E6 
for $p=1,10,50,100$. 
For $p=1, 10$, in all cases of E1-E6, the Wilcoxon signed-rank test at the 5\% significance level 
did not reject the null hypothesis. 
For $p=50, 100$, the Wilcoxon signed-rank test at the 5\% significance level 
did not reject the null hypothesis 
with respect to E1, E2, E3, E5 and E6, 
while it rejected the null hypothesis with respect to E4. 
From those statistical tests, we observe the followings.
When the time horizon is relatively short (i.e., for $p=1, 10$), 
all of E1-E6 have desirable statistical properties regarding bias. 
However, as the time horizon extends to $p=50, 100$, E4 no longer has desirable properties regarding bias. 
We therefore conjecture that all portfolios E1--E6 excluding E4 possess desirable statistical properties regarding bias 
as tracking portfolios. 
For reference, we provide the sample mean of residuals $\varepsilon^{p}_{t}$ $(p=1,10,50,100)$ across all data points
in Table \ref{sample-mean-TOPIX}, where items shaded in light blue 
indicate that the null hypothesis of the Wilcoxon signed-rank test is rejected at the 5\% significance level.

\begin{table}
  \begin{center}
    \caption{Sample variance of residual for cumulative return (TOPIX) ($\times 10^{-5}$)}
    \label{sample-variance-TOPIX}
    \begin{tabular}{crrrrrr} 
      \toprule 
       & E1 & E2 & E3 & E4 & E5 & E6 \\
      \midrule 
      $\cellcolor{orange!10}p=1$ & $0.568$ & $0.586$ & $0.627$ & $0.832$ 
      & $0.717$ & $0.541$ \\
      $\cellcolor{orange!10}p=10$ & $5.56$ & $5.30$ & $5.62$ & $7.07$
      & $6.29$ & $5.21$ \\
      $p=50$ & $24.5$ & $26.8$ & $31.6$ & $\cellcolor{blue!10} \textcolor{gray}{33.0}$
      & $30.3$ & $28.6$ \\
      $p=100$ & $53.9$ 
      & $56.8$ & $72.5$ 
      & $\cellcolor{blue!10} \textcolor{gray}{75.5}$
      & $72.1$ & $62.9$ \\
      
      \bottomrule 
    \end{tabular}
        
  \end{center}
\end{table}

We next focus on the variances of residuals $\varepsilon^{p}_{t}$ 
as estimates for the tracking error of the tracking portfolios E1--E6.
For $p=1, 10$, 
we apply the Levene test to the residuals $\varepsilon^{p}_{t}$ of E1--E6, 
which passed the Wilcoxon signed-rank test for $p=1, 10$. 
For $p=1, 10$, the Levene test at the 5\% significance level 
rejected the null hypothesis that all the input samples are 
from populations with equal variances. 

For $p=50, 100$, we apply the Levene test to the residuals $\varepsilon^{p}_{t}$ of E1, E2, E3, E5 and E6, 
which passed the Wilcoxon signed-rank test for $p=50, 100$.
The Levene test at the 5\% significance level 
did not reject the null hypothesis. 
For reference, we provide the sample variances of residuals $\varepsilon^{p}_{t}$ 
$(p=1,10,50,100)$ across all data points in Table \ref{sample-variance-TOPIX}, 
where items shaded in light blue  
indicate that the Levene test was not applied (because  
the null hypothesis of the Wilcoxon signed-rank test at the 5\% significance level 
was rejected) and 
lines with light orange shaded $p$'s indicate that the null hypothesis of the Levene test at the 5\% 
significance level was rejected. 
In Table \ref{sample-variance-TOPIX}, 
we observe that for $p=1, 10$, where the null hypothesis of the Levene test was rejected,
E6 consistently provides the most desirable property regarding variance among 
E1--E6. 
We therefore conjecture that as sparse tracking portfolio, 
E6 possesses the most desirable statistical properties regarding bias and variance 
among E1--E6.

\begin{table}
  \begin{center}
    \caption{Sample mean of absolute value of residuals (TOPIX) ($\times 10^{-2}$)}
    \label{sample-mean-TOPIX-year}
    \begin{tabular}{crrrrrr} 
    \toprule 
     & E1 & E2 & E3 & E4 & E5 & E6 \\
    \midrule 
    1 year & $3.49$ & $3.19$ & $3.73$ & $3.29$ 
    & $3.62$ & $2.85$ \\
    2 years & $5.47$ & $4.18$ & $4.75$ & $4.72$
    & $4.32$ & $3.08$ \\
    \bottomrule 
    \end{tabular}
      
  \end{center}
  \end{table}

  \begin{table}
    \begin{center}
      \caption{Sample variance of absolute value of residuals (TOPIX) ($\times 10^{-4}$)}
      \label{sample-variance-TOPIX-year}
      \begin{tabular}{crrrrrr} 
      \toprule 
       & E1 & E2 & E3 & E4 & E5 & E6 \\
      \midrule 
      1 year & $3.97$ & $4.70$ & $6.35$ & $7.96$ 
      & $7.30$ & $3.60$ \\
      2 years & $12.5$ & $9.92$ & $11.2$ & $16.9$
      & $10.6$ & $4.67$ \\
      \bottomrule 
      \end{tabular}
        
    \end{center}
    \end{table}

\begin{figure}
  \centering
  \includegraphics[width=7.2in]{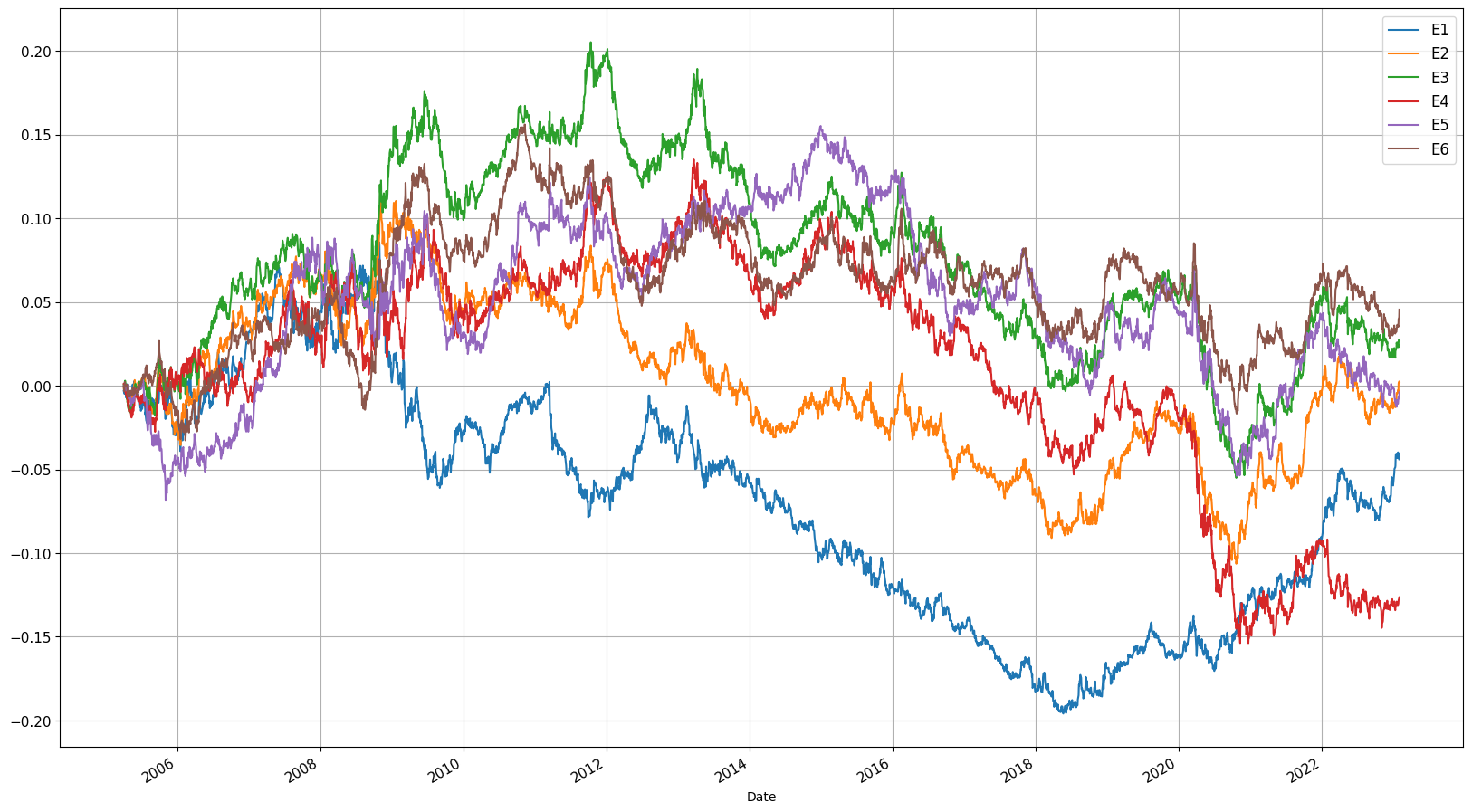}
  \caption{Residuals $\varepsilon^{p}_{1}$ for cumulative return as a function of $p$ (TOPIX)}
  \label{fig-TOPIX}
\end{figure}

Finally, for reference, 
we observe the sample mean and sample variance of absolute value of  the residuals for longer time horizon, 
specifically, for one year and two years, 
where we did not apply the statistical tests due to the limitation of the available data size. 
Tables \ref{sample-mean-TOPIX-year} and \ref{sample-variance-TOPIX-year} 
display the sample mean and sample variance of absolute value of the residuals, respectively,  
for one year and two years across all data points. 
In Tables \ref{sample-mean-TOPIX-year} and \ref{sample-variance-TOPIX-year}, we observe that 
E6 is superior to E1--E5 in terms of the sample mean and sample variance of absolute value of the residuals. 
Also, for reference, 
we illustrate the residuals $\varepsilon^{p}_{t}$ of tracking portfolios E1--E6 
by fixing the beginning date $t=1$
as a function of time horizon $p$ $(p=1,\ldots, 4366)$ in Figure \ref{fig-TOPIX}. 
In Figure \ref{fig-TOPIX}, a smaller area formed by the curve and the line $y=0$ implies a better tracking performance. 
In Figure \ref{fig-TOPIX}, we observe that E1 and E3 exhibit 
larger areas formed by their curve and the line $y=0$ and indicate the 
presence of periods with a considerable positive or negative biases. 
On the other hand, compared to E1 and E3, E6  
is stable and remain close to the line $y=0$. 
We see that $\max_{p} | \varepsilon^{p}_{1} | \approx 0.16$ in E6, 
whereas  $\max_{p} | \varepsilon^{p}_{1} | \approx 0.20$ in E1 and E3.


Next, we examine the tracking performance of the tracking portfolios E1--E6 
for S\&P500. 
Similar to the case of TOPIX, 
we introduce the residuals $\varepsilon^{p}_{t}$ for cumulative return with time horizon $p$. 
We then examine the residuals $\varepsilon^{p}_{t}$ of E1--E6 for $p=1,10,50,100$. 
We apply the similar statistical test procedure as in the case of TOPIX 
to examine the residuals $\varepsilon^{p}_{t}$ of E1--E6 in the case of S\&P500.

\begin{table}
  \begin{center}
    \caption{Sample mean of residual for cumulative return (S\&P500) ($\times 10^{-5}$)}
    \label{sample-mean-SP500}
    \begin{tabular}{crrrrrr} 
    \toprule 
     & E1 & E2 & E3 & E4 & E5 & E6 \\
    \midrule 
    $p=1$ & $-0.326$ & $5.92$ & $5.72$ & $1.82$ 
    & $-5.65$ & $0.948$ \\
    $p=10$ & $-6.75$ 
    & $57.3$ 
    & $56.8$ 
    & $\cellcolor{blue!10} \textcolor{gray}{14.7}$
    & $\cellcolor{blue!10} \textcolor{gray}{-57.0}$ 
    & $9.79$ \\
    $p=50$ & $38.2$ & $\cellcolor{blue!10} \textcolor{gray}{322}$ 
    & $317$ & $59.8$    
    & $\cellcolor{blue!10} \textcolor{gray}{-293}$ & $78.8$ \\
    $p=100$ & $39.6$ 
    & $\cellcolor{blue!10} \textcolor{gray}{666}$ & $\cellcolor{blue!10} \textcolor{gray}{631}$  
    & $53.9$
    & $\cellcolor{blue!10} \textcolor{gray}{-683}$ 
    & $156$ \\
    \bottomrule 
    \end{tabular}
      
  \end{center}
  \end{table}

  \begin{table}
    \begin{center}
      \caption{Sample variance of residual for cumulative return (S\&P500) ($\times 10^{-5}$)}
      \label{sample-variance-SP500}
      \begin{tabular}{crrrrrr} 
      \toprule 
       & E1 & E2 & E3 & E4 & E5 & E6 \\
      \midrule 
      $\cellcolor{orange!10} p=1$ & $0.457$ & $0.335$ & $0.400$ & $0.999$ 
      & $0.670$ & $0.311$ \\
      $\cellcolor{orange!10} p=10$ & $4.59$ 
      & $2.96$ 
      & $3.66$ 
      & $\cellcolor{blue!10} \textcolor{gray}{9.35}$
      & $\cellcolor{blue!10} \textcolor{gray}{6.65}$ 
      & $3.06$ \\
      $\cellcolor{orange!10} p=50$ & $26.6$ & $\cellcolor{blue!10} \textcolor{gray}{11.3}$ 
      & $16.7$ & $48.6$    
      & $\cellcolor{blue!10} \textcolor{gray}{27.4}$ & $13.5$ \\
      $\cellcolor{orange!10} p=100$ & $69.5$ 
      & $\cellcolor{blue!10} \textcolor{gray}{21.6}$ & $\cellcolor{blue!10} \textcolor{gray}{28.7}$  
      & $106$
      & $\cellcolor{blue!10} \textcolor{gray}{57.2}$ 
      & $24.1$ \\
      \bottomrule 
      \end{tabular}
      
  \end{center}
  \end{table}

Similar to the case of TOPIX, for all the cases of $p=1,10,50, 100$, 
the Shapiro-Wilk test indicated that the residuals $\varepsilon^{p}_{t}$ of E1--E6 did not all follow 
normal distributions at the 5\% significance level. 
To examine the biases of the residuals,
we thus applied the Wilcoxon signed-rank test 
to the residuals $\varepsilon^{p}_{t}$ of E1--E6 for $p=1,10,50,100$. 
We provide the results of the Wilcoxon signed-rank test at the 5\% significance level 
and the sample means of residuals $\varepsilon^{p}_{t}$ 
$(p=1,10,50,100)$ in Table \ref{sample-mean-SP500}, 
where item shaded in light blue indicate that the null hypothesis 
of the Wilcoxon signed-rank test at the 5\% significance level was rejected. 
We also provide the results of the Levene test at the 5\% significance level 
and the sample variances of residuals $\varepsilon^{p}_{t}$ 
$(p=1,10,50,100)$ in Table \ref{sample-variance-SP500}, where items shaded in light blue indicate that 
the Levene test was not applied (because  
the null hypothesis of the Wilcoxon signed-rank test at the 5\% significance level 
was rejected)
and lines with light orange shaded $p$'s indicate that the null hypothesis of the Levene test at the 5\% 
significance level was rejected.

In Tables \ref{sample-mean-SP500}, we observe that 
E1 and E6 exhibit desirable statistical properties regarding bias, 
because the null hypothesis of the Wilcoxon signed-rank test was not rejected for all $p=1,10,50,100$.
In Table \ref{sample-variance-SP500}, 
we observe that for $p=1, 10, 50, 100$, where the null hypothesis of the Levene test was rejected,
E6 consistently provides smaller sample variances than E1.  
We therefore conjecture that as sparse tracking portfolio, 
E6 possesses the most desirable statistical properties regarding bias and variance 
among E1--E6.

\begin{table}
  \begin{center}
    \caption{Sample mean of absolute value of residuals (S\&P500) ($\times 10^{-2}$)}
    \label{sample-mean-SP500-year}
    \begin{tabular}{crrrrrr} 
    \toprule 
     & E1 & E2 & E3 & E4 & E5 & E6 \\
    \midrule 
    1 year & $4.08$ & $2.71$ & $2.56$ & $4.35$ 
    & $3.09$ & $2.00$ \\
    2 years & $6.86$ & $5.41$ & $4.61$ & $7.27$
    & $5.79$ & $3.43$ \\
    \bottomrule 
    \end{tabular}
      
  \end{center}
  \end{table}

  \begin{table}
    \begin{center}
      \caption{Sample variance of absolute value of residuals (S\&P500) ($\times 10^{-4}$)}
      \label{sample-variance-SP500-year}
      \begin{tabular}{crrrrrr} 
      \toprule 
       & E1 & E2 & E3 & E4 & E5 & E6 \\
      \midrule 
      1 year & $14.9$ & $2.07$ & $2.87$ & $12.3$ 
      & $9.06$ & $1.92$ \\
      2 years & $37.1$ & $8.95$ & $9.78$ & $28.4$
      & $33.6$ & $4.44$ \\
      \bottomrule 
      \end{tabular}
        
    \end{center}
    \end{table}

    \begin{figure}
      \centering
      \includegraphics[width=7.2in]{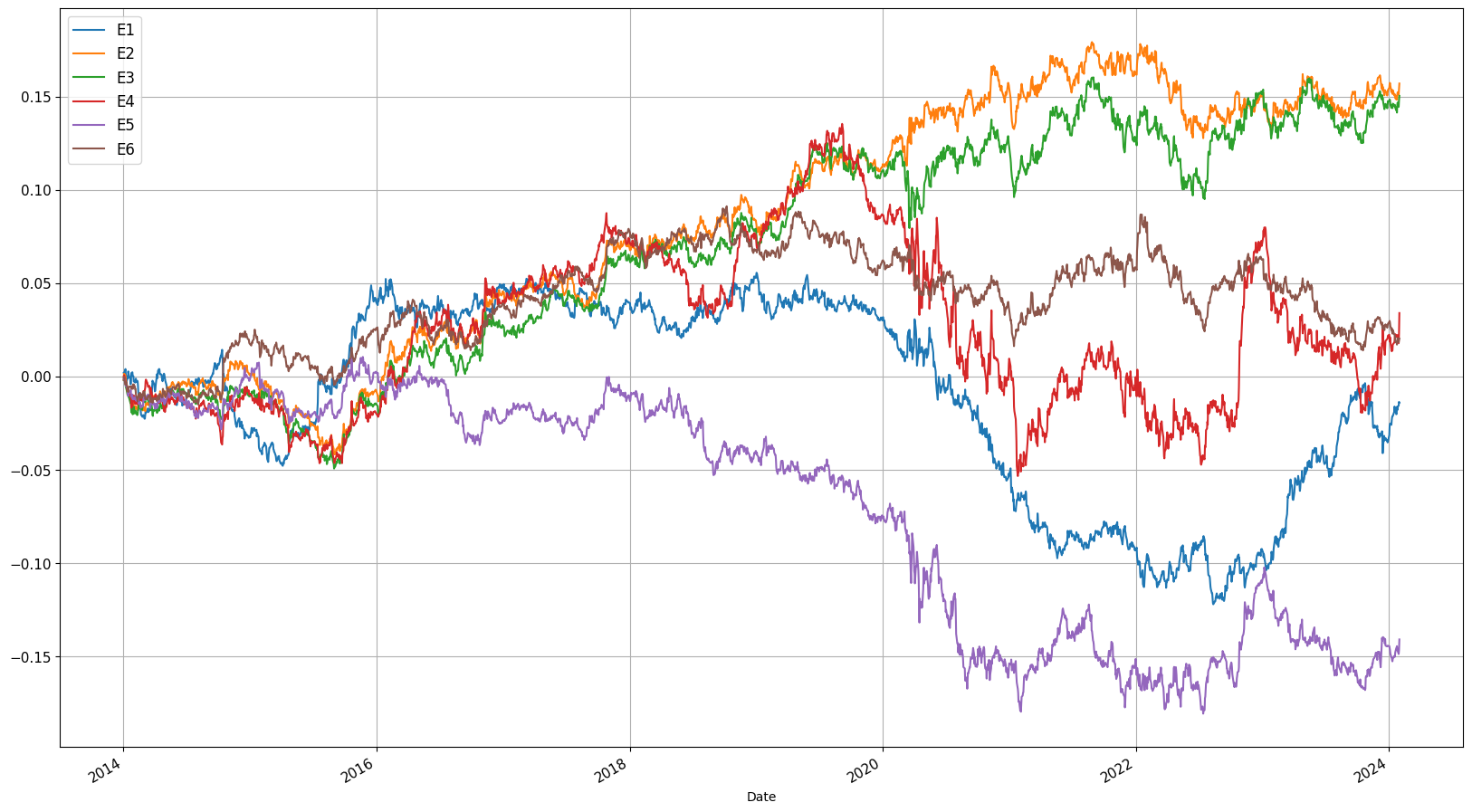}
      \caption{Residuals $\varepsilon^{p}_{1}$ for cumulative return as a function of $p$ (S\&P500)}
      \label{fig-SP500}
    \end{figure}
    
    Finally, for reference, 
    we observe the sample mean and sample variance of absolute value of the residuals for longer time horizon, 
    specifically, for one year and two years, 
    where we did not apply the statistical tests due to the limitation of the available data size. 
    Tables \ref{sample-mean-SP500-year} and \ref{sample-variance-SP500-year} 
    display the sample mean and sample variance of absolute value of the residuals, respectively,  
    for one year and two years across all data points. 
    In Tables \ref{sample-mean-SP500-year} and \ref{sample-variance-SP500-year}, we observe that 
E6 is superior to E1--E5 in terms of the sample mean and sample variance of absolute value of the residuals. 
    Also, for reference, 
    we illustrate the residuals $\varepsilon^{p}_{t}$ of tracking portfolios E1--E6 
    by fixing the beginning date $t=1$
    as a function of time horizon $p$ $(p=1,\ldots, 2537)$ in Figure \ref{fig-SP500}. 
    In Figure \ref{fig-SP500}, a smaller area formed by the curve and the line $y=0$ implies a better tracking performance. 
    In Figure \ref{fig-SP500}, we observe that E2 and E5 exhibit 
    larger areas formed by their curve and the line $y=0$ and indicate the 
    presence of periods with a considerable positive or negative biases. 
    On the other hand, compared to E2 and E5, E6  
    is stable and remain close to the line $y=0$. 
    We see that $\max_{p} | \varepsilon^{p}_{1} | \approx 0.09$ in E6, 
    whereas  $\max_{p} | \varepsilon^{p}_{1} | \approx 0.18$ in E2 and E5.


\section{Conclusion}
\label{sec:Conclusion}

In this study, we propose formulations to select assets for portfolios tracking stock index. 
The proposed formulations are described as combinatorial optimization problems. 
Our formulations consider the impact of  assets with large MC on the stock index, 
centrality and dissimilarity of selected assets. 
They include MC top-tier selection and CBS as special cases, and 
they can yield various mixture selections of these selection by adjusting (hyper) parameters. 
In addition to single-stage selection method, we propose multi-stage selection method, 
where asset selection is performed multiple times with different values of 
parameters governing centrality and dissimilarity   

We apply our formulations to construct tracking portfolios of TOPIX and those of S\&P500, 
and provide some numerical examples for their resulting tracking portfolios. 
The numerical examples exhibit that by adjusting the parameters included in the proposed formulations, 
it is possible to construct a sparse tracking portfolio that outperforms the tracking performance of 
existing MC top-tier or CBS with single-stage selection.

\section*{Acknowledgment}

The authors would like to thank Mr. Hajime Ota of NTT DATA for organizing the meetings that led to the development of this work.

\bibliographystyle{apalike}
\bibliography{tracking_portfolio}

\begin{thebibliography}{}

\bibitem[Alipour et~al., 2016]{alipour2016quantum}
Alipour, E., Adolphs, C., Zaribafiyan, A., and Rounds, M. (2016).
\newblock Quantum-inspired hierarchical risk parity.
\newblock {\em White paper, 1Qbit}.

\bibitem[Bauckhage et~al., 2019]{bauckhage2019qubo}
Bauckhage, C., Piatkowski, N., Sifa, R., Hecker, D., and Wrobel, S. (2019).
\newblock A qubo formulation of the k-medoids problem.
\newblock In {\em LWDA}, pages 54--63.

\bibitem[Beasley et~al., 2003]{beasley2003evolutionary}
Beasley, J.~E., Meade, N., and Chang, T.-J. (2003).
\newblock An evolutionary heuristic for the index tracking problem.
\newblock {\em European Journal of Operational Research}, 148(3):621--643.

\bibitem[Benidis et~al., 2017]{benidis2017sparse}
Benidis, K., Feng, Y., and Palomar, D.~P. (2017).
\newblock Sparse portfolios for high-dimensional financial index tracking.
\newblock {\em IEEE Transactions on signal processing}, 66(1):155--170.

\bibitem[Benidis et~al., 2018]{benidis2018optimization}
Benidis, K., Feng, Y., Palomar, D.~P., et~al. (2018).
\newblock Optimization methods for financial index tracking: From theory to practice.
\newblock {\em Foundations and Trends in Optimization}, 3(3):171--279.

\bibitem[Dose and Cincotti, 2005]{dose2005clustering}
Dose, C. and Cincotti, S. (2005).
\newblock Clustering of financial time series with application to index and enhanced index tracking portfolio.
\newblock {\em Physica A: Statistical Mechanics and its Applications}, 355(1):145--151.

\bibitem[Fastrich et~al., 2015]{fastrich2015constructing}
Fastrich, B., Paterlini, S., and Winker, P. (2015).
\newblock Constructing optimal sparse portfolios using regularization methods.
\newblock {\em Computational Management Science}, 12(3):417--434.

\bibitem[Feng et~al., 2016]{feng2016signal}
Feng, Y., Palomar, D.~P., et~al. (2016).
\newblock A signal processing perspective on financial engineering.
\newblock {\em Foundations and Trends in Signal Processing}, 9(1--2):1--231.

\bibitem[Hong et~al., 2021]{hong2021market}
Hong, S.~W., Miasnikof, P., Kwon, R., and Lawryshyn, Y. (2021).
\newblock Market graph clustering via qubo and digital annealing.
\newblock {\em Journal of Risk and Financial Management}, 14(1):34.

\bibitem[Jansen and Van~Dijk, 2002]{jansen2002optimal}
Jansen, R. and Van~Dijk, R. (2002).
\newblock Optimal benchmark tracking with small portfolios.
\newblock {\em Journal of Portfolio Management}, 28(2):33.

\bibitem[Rosenberg et~al., 2016]{rosenberg2016solving}
Rosenberg, G., Haghnegahdar, P., Goddard, P., Carr, P., Wu, K., and De~Prado, M.~L. (2016).
\newblock Solving the optimal trading trajectory problem using a quantum annealer.
\newblock {\em IEEE Journal of Selected Topics in Signal Processing}, 10(6):1053--1060.

\bibitem[Sakurai et~al., 2021]{sakurai2021correlation}
Sakurai, Y., Yuki, Y., Katsuki, R., Yazane, T., and Ishizaki, F. (2021).
\newblock Correlation diversified passive portfolio strategy based on permutation of assets.
\newblock {\em Journal of Investment Strategies}, 10(2).

\bibitem[Stanley and Mantegna, 2000]{stanley2000introduction}
Stanley, H.~E. and Mantegna, R.~N. (2000).
\newblock {\em An introduction to econophysics}.
\newblock Cambridge University Press, Cambridge.

\bibitem[Tatsumura et~al., 2023]{tatsumura2023real}
Tatsumura, K., Hidaka, R., Nakayama, J., Kashimata, T., and Yamasaki, M. (2023).
\newblock Real-time trading system based on selections of potentially profitable, uncorrelated, and balanced stocks by np-hard combinatorial optimization.
\newblock {\em arXiv preprint arXiv:2307.06339}.

\bibitem[Varsei et~al., 2013]{varsei2013heuristic}
Varsei, M., Shams, N., Fahimnia, B., and Yazdanpanah, A. (2013).
\newblock A heuristic approach to the index tracking problem: a case study of the tehran exchange price index.
\newblock {\em Asian academy of management Journal}, 18(1):19.

\end{thebibliography}

\end{document}